\documentclass[a4paper,11pt]{article}
\usepackage{pos}
\usepackage{graphicx}
\usepackage[colorinlistoftodos]{todonotes}
\usepackage{booktabs}
\usepackage{caption}
\usepackage{subcaption} 
\usepackage{multirow} 
\usepackage{multicol}
\usepackage{algorithm}
\usepackage[noend]{algpseudocode}
\usepackage[section]{placeins}
\usepackage[normalem]{ulem}
\usepackage{todonotes}

\usepackage{bm}             
\usepackage{physics}
\usepackage{dsfont}         
\usepackage{amsopn}         
\usepackage{mathrsfs}       
\usepackage{xfrac}          
\usepackage{bbold}          
\usepackage{slashed}        
\usepackage{mathtools}      

\usepackage{hyperref}

 \newcommand{\be}{\begin{equation}}
\newcommand{\ee}{\end{equation}}
\newcommand{\ba}{\begin{align}}
\newcommand{\ea}{\end{align}}
\newcommand{\baa}{\begin{array}}
\newcommand{\eaa}{\end{array}}
\newcommand{\bea}{\begin{eqnarray}}
\newcommand{\eea}{\end{eqnarray}}

\newcommand{\tl}{\tilde{l}}

\newcommand{\one}{\mathbb{1}}
\title{Adjoint fermions at large-$N_c$ on the lattice}

\author*[a,b]{P. Butti}
\author[b]{M. Garc\'ia P\'erez}
\author[a,b]{A. Gonz\'alez-Arroyo}
\author[c,d]{K. I. Ishikawa}
\author[d]{M. Okawa}

\affiliation[a]{Departamento de F\'{\i}sica Te\'orica, M\'odulo 15, Universidad Aut\'onoma de Madrid,
Cantoblanco, E-28045, Madrid, Spain}

\affiliation[b]{Instituto de F\'{\i}sica Te\'orica UAM-CSIC,
  Calle Nicol\'as Cabrera 13-15, Universidad Aut\'onoma de Madrid, Cantoblanco, E-28049, Madrid, Spain}
 
\affiliation[c]{Core of Research for the Energetic Universe, Graduate School of Advanced Science and Engineering, Hiroshima University, Higashi-Hiroshima, Hiroshima 739-8526, Japan}

\affiliation[d]{Graduate School of Advanced Science and Engineering, Hiroshima University, Higashi-Hiroshima, Hiroshima 739-8526, Japan}

\emailAdd{pietro.butti@uam.es}
\emailAdd{margarita.garcia@csic.es}
\emailAdd{antonio.gonzalez-arroyo@uam.es}
\emailAdd{ishikawa@theo.phys.sci.hiroshima-u.ac.jp}
\emailAdd{okawa@hiroshima-u.ac.jp}
\abstract{
Lattice simulations of Yang-Mills theories coupled with $N_f$ flavours of fermions in the adjoint representation provide a way to probe the non-perturbative regime of a plethora of different physical scenarios, such as Supersymmetric Yang-Mills theory to BSM models. We are interested in the large-$N_c$ limit of these theories, for which standard lattice techniques are limited by high-computational costs. Our approach makes use of the well-established twisted volume reduction, which allows one to simulate these theories on a $1^4$ lattice. In this talk, we are going to present a detailed study of the scale setting of these theories, performed with Wilson flow techniques, but endowed with a procedure that allowed us to reduce finite-volume (finite $N_c$) systematic effects. We will apply this procedure to our configurations for $N_f=0,\frac{1}{2},1,2$ and analyze  the dependence of the scale on the coupling, the adjoint fermion mass and the number of flavours. For the cases in which enough couplings are available we compared the resulting  $\beta$-function with the predictions of perturbation theory, finding very good agreement.
}

\FullConference{%
 The 39th International Symposium on Lattice Field Theory, LATTICE2022
  6th-13th August, 2022
 Bonn, Germany  
}

\begin{document}

\begin{flushright}
    \vspace*{-8em}
	IFT-UAM/CSIC-22-146 HUPD-2214
    \vspace*{4em}
\end{flushright}
\maketitle

\section{Introduction}
$SU(N_c)$ gauge theory coupled with several flavours of fermions in the adjoint representation is a theoretically appealing scenario of high relevance for several different frameworks. As an example, the case of $N_f=\frac{1}{2}$ (one adjoint Majorana fermion), corresponds to $\mathcal{N}=1$ SUSY Yang-Mills which shares several non-trivial properties with QCD, while maintaining minimal supersymmetry. On the other hand, the case of $N_f=1$ (one adjoint Dirac fermion) is of high relevance for BSM models and the study of condensed matter systems, and that of $N_f=2$ (two adjoint Dirac fermions) is a valuable test case for conformality. The most interesting regime of these theories is at strong coupling where they manifest many non-trivial phenomena, depending on their matter content.

We are interested in the study of the large-$N_c$ limit of these theories and our approach makes use of volume reduction using twisted boundary conditions \cite{Gonzalez-Arroyo:1982hyq,Gonzalez-Arroyo:2010omx}. This methodology is a well-established technique that allows one to simulate gauge theory on a single site lattice and it has already been applied in the past in the context of pure Yang-Mills theory~\cite{Gonzalez-Arroyo:2012euf,Perez:2020vbn}, with dynamical $N_f=1,2$ adjoint Dirac fermions \cite{Gonzalez-Arroyo:2013bta,GarciaPerez:2015rda} and recently for $N_f=\frac{1}{2}$ \cite{Butti:2022sgy}.

In this talk, we will first review general concepts of twisted volume reduction. Next we will go through the methodology we developed to perform scale setting. Our setup allows us to have control over the systematic effects related to finite-volume (finite-$N_c$), including the case of small fermion masses. We will then present the results of  applying  the method to large $N_c$ configurations involving various flavours of dynamical adjoint fermions. This will allow us to analyze the dependence of the scale on the various parameters including the quark masses. Interesting conclusions will then be drawn concerning scaling and the comparison of our results with the perturbative $\beta$-function.

\subsection{Twisted volume reduction at large-$N_c$ limit}
Our model consists of a standard $SU(N_c)$ gauge theory discretized on a $1^4$ lattice with twisted boundary conditions. After a trivial change of variables, the resulting Wilson action takes the following form
\begin{equation}\label{eqn:tekaction}
    S_\text{TEK} = bN_c \sum_{\mu\neq\nu}\tr[\one - z_{\mu\nu} U_\mu U_\nu U_\mu^\dagger U_\nu^\dagger],
\end{equation}
where $U_\mu$ are $SU(N_c)$ matrices, $b=\frac{1}{\lambda}$ is the inverse of 't Hooft coupling $\lambda=g^2 N_c$, while $z_{\mu\nu}=e^{i\frac{2\pi k}{\sqrt{N_c}}}$, $z_{\nu\mu}=z_{\mu\nu}^*$, is a numerical factor which encodes the twist ($k$ is an integer number selected coprime with $\sqrt{N_c}$). This particular choice of the twist factor is called \textit{symmetric twist}. The vacuum configurations for the TEK action \eqref{eqn:tekaction} are called \textit{twist eaters} and they can be defined as the solution of the twist equation
\begin{equation}
    \Gamma_\mu \Gamma_\nu = z_{\nu\mu} \Gamma_\nu \Gamma_\mu.
\end{equation}
In perturbation theory, one can see easily that momenta circulating in diagrams are quantized as integer multiples of $\frac{2\pi}{\sqrt{N_c}}$, which is an indication that our reduced models describe the physics of an effective lattice whose size is $V=(\sqrt{N_c})^4$. Volume independence states that, in the large-$N_c$ limit, expectation values of gauge-invariant single-trace observables recover the infinite volume limit of the same observables calculated in the standard Wilson-like theory \cite{Gonzalez-Arroyo:1982hyq, Gonzalez-Arroyo:2010omx}.

Given the adjoint nature of our fermions, we can apply the same twisted reduction procedure to quark fields \cite{Gonzalez-Arroyo:2013bta,GarciaPerez:2015rda,Butti:2022sgy}. We are then left with the twisted reduced version of the Wilson-Dirac operator
\begin{equation}
    D_w = \one - \kappa\sum_{\mu} \qty[(\one-\gamma_\mu)U_\mu^\text{adj} + (\one+\gamma_\mu){U_\mu^\text{adj}}^\dagger],
\end{equation}
where $U_\mu^\text{adj}$ is the link field in the adjoint representation. Details about the numerical implementation and further references can be found in~\cite{Butti:2022sgy}.

\section{Scale setting with Wilson flow}
We consider the evolution of gauge field configurations driven by the Wilson flow equation $\partial_t A_\mu(x,t) = D_\nu G_{\mu\nu}(x,t)$, $t$ being the flow time. A common observable is the \text{flowed energy density} $E(t) = \expval{\tr \qty(G_{\mu\nu}(x,t)G_{\mu\nu}(x,t))}$, which, on a single-site twisted lattice, can be written in the following way
\begin{equation}\label{eqn:symme}
    E = -\frac{1}{128}\sum_{\mu\neq\nu} \tr[z_{\nu\mu}(
    U_\nu U_\mu U_\nu^\dagger U_\mu^\dagger +
    U_\mu U_\nu^\dagger U_\mu^\dagger U_\nu +
    U_\nu^\dagger U_\mu^\dagger U_\nu U_\mu + 
    U_\mu^\dagger U_\nu U_\mu U_\nu^\dagger  - \text{h.c.}
    )]^2.
\end{equation}
The lattice version of the flow equations are integrated with respect to the lattice flow time $T=\frac{t}{a^2}$. 
In the following, we will consider the dimensionless version of the flowed energy density normalized to the number of colours, $\Phi(T) = \frac{\expval{T^2E(T)}}{N_c}$. 
\begin{figure}[t]
    \centering
    \includegraphics[scale=0.3]{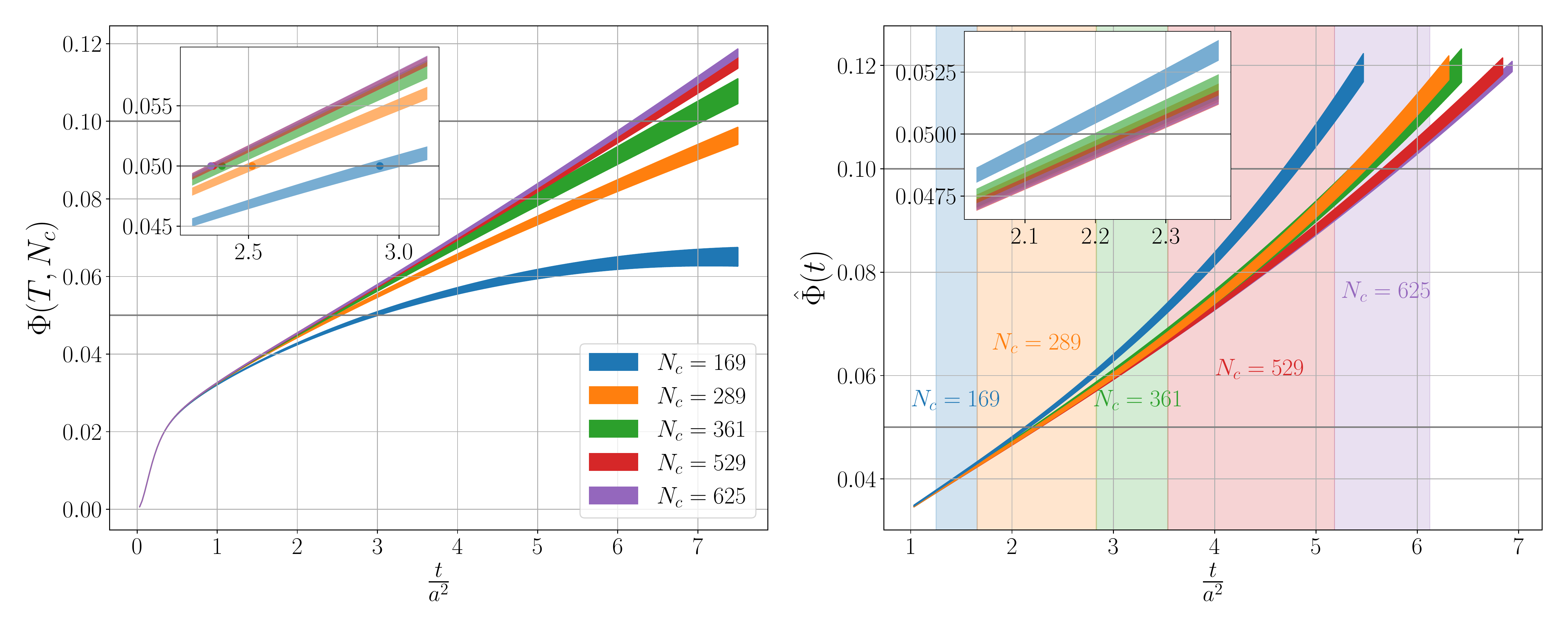}
    \caption{\textit{Left-hand side}: Flowed dimensionless energy density for pure Yang-Mills for different numbers of colours at $b=0.37$. \textit{Right-hand side}: Same flow curves with the norm correction applied. The scaling windows in Eq.~\eqref{eqn:scalingw} are represented with a vertical stripe of the same colour as the flow curve they are associated with. They all start at $T=1.25$ but it is only depicted the ones with the smallest $N_c$ not to make the colours overlap.}
    \label{fig:flow}
\end{figure}
After the flowed energy density is calculated, a scale can be defined  by choosing a reference value $s$ and then solving the following implicit equation
\begin{equation}\label{eqn:scaleeq}
    \eval{\Phi(T)}_{T=T(s)} \equiv \frac{1}{N_c}\eval{ \expval{T^2 E(T)}}_{T=T(s)} = s.
\end{equation}
The standard choice for the reference scale in QCD simulations is $s=s_0\equiv0.1$\footnote{Note that in standard lattice QCD simulations the dimensionless flowed energy density is not divided by the number of colours as we do in Eq.~\eqref{eqn:scaleeq}, therefore our $T_0$ for $s=0.1$ corresponds to $s=0.3$ in QCD simulations.}, whose relative scale will be called $T_0$, but also other values of $s$ can be adopted. In particular, the choice $s=s_1\equiv 0.05$ produces a scale $T_1$ that results to be less affected by finite-volume effects while still being safe from lattice artefacts. In the left side of Fig.~\ref{fig:flow} we display the flow curves for several values of $N_c$ for the case of pure Yang-Mills ($N_f=0$) at $b=0.37$, where the two reference scales $s_0=0.1$ and $s_1=0.05$ are depicted with corresponding horizontal grey lines. As it is evident from the magnification in the plot, each curve corresponding to a different $N_c$ intersects the line corresponding to $s_1$ at a different value of the lattice flow time $T$. In general, one could attempt an extrapolation of the scale to $N_c\rightarrow{\infty}$, but several values of $N_c$ would be needed to have good control over the quality of the extrapolation. Since in most of our simulations we just have 2 or 3 values of $N_c$ available, we adopted another strategy. In \cite{Butti:2022sgy} we analyzed in detail the source of these $N_c$ dependence effects and developed a method that we applied to the case of $N_f=\frac{1}{2}$. In the following subsection, we will give a general indication of how this method is implemented, referring to \cite{Butti:2022sgy} for further technical details.

\subsection{Analysis of finite size effects}
In our set-up, the number of colours plays the role of the volume of an effective lattice through the relation $V=(\sqrt{N_c})^4$, therefore these finite-$N_c$ effects can be treated analogously to finite-size effects on standard lattice simulations. The main observation is that, as analyzed in detail in \cite{GarciaPerez:2014azn}, lattice perturbation theory at LO states that the $N_c$ dependence of $\Phi(T,N_c)$ is given by an overall coefficient that  can be computed analytically $\mathcal{N}_L(T,N_c)$ and whose expression can be found in Eq.~(B.10) in Ref.~\cite{Butti:2022sgy}. In light of this, one can employ a modified version of $\Phi(T)$, given by
\begin{equation}
    \hat{\Phi}(T) = \frac{\frac{3}{128\pi^4}}{\mathcal{N}_L(T,N_c)}\Phi(T,N_c)
\end{equation}
where the number $\frac{3}{128\pi^4}$ is the analytic value of the overall coefficient in the continuum and in the $N_c\rightarrow\infty$ limit, as known from \cite{Luscher:2010iy}. This modification takes into account both finite-$N_c$ effects and lattice artefacts to LO in perturbation theory. This method is expected to work within a \textit{scaling window}, whose lower bound is determined by limiting the effects of lattice artefacts and the higher bound by the validity of PT to LO and remnant finite-size effects. We define this scaling window to be
\begin{equation}\label{eqn:scalingw}
    T\in\qty[1.25,\gamma^2\frac{N}{8}]\text{ with }\gamma\sim 0.3
\end{equation}
where we empirically verify that curves do not suffer from the influence of lattice artefacts nor remnant finite volume effects. For the previously considered example of pure Yang-Mills theory at $b=0.37$, the effects of the corrections are depicted on the right-hand side of  Fig.~\ref{fig:flow}, where each colour in the plot corresponds to a different $N_c$ and the different scaling windows are represented as coloured vertical bands.\footnote{Each coloured band starts at $T=1.25$. Whenever two or more bands overlap we only depict the one with the lowest $N_c$.}  It is visible by eye that inside each scaling window all the different flow curves collapse.
An important observation is that the reference value $s_1$ falls inside the scaling window of all the cases (except for $N_c=169$ ), therefore $\sqrt{8T_1}$ can be safely estimated through an interpolation. This is not true for $s_0=0.1$, which falls out of the scaling windows (except for $N_c=625$), and therefore the scale has to be calculated with an extrapolation done by  fitting our data for different values of $b$ to a single universal curve, whose argument is $T/T_1(b,N_c)$. For this reason, we preferred to perform the scale setting of all our configurations using $\sqrt{8T_1}$ instead of $\sqrt{8T_0}$, also having in mind that for the majority of ensembles for cases of $N_f=\frac{1}{2},1,2$ we have at our disposal configurations for only {\em smaller} values of $N_c=289,361$. 

The numerical values for $\sqrt{8T_0}$ and $\sqrt{8T_1}$ for different values of $N_c$ are reported in Tab.~\ref{tab:scales_YM}. With the exception of $N_c=169$, all scales result to be compatible within errors.
\begin{table}[]
    \centering
    \begin{tabular}{lll}
    \toprule
        $N_c$ & $\frac{\sqrt{8t_1}}{a}$ & $\frac{\sqrt{8t_0}}{a}$ \\
    \midrule
        169 & 4.141(73) & 6.5(1.6) \\
        289 & 4.237(69) & 6.78(21) \\
        361 & 4.222(88) & 6.69(15) \\
        529 & 4.250(56) & 6.780(41)  \\
        625 & 4.256(32) & 6.800(19)  \\
    \bottomrule
    \end{tabular}
    \caption{Values of the Wilson-flow scales for pure Yang-Mills theory at $b=0.37$ for different value of $N_c$. $\sqrt{T_1}$ is estimated through an interpolation, $\sqrt{8T_0}$ through extrapolation.}
    \label{tab:scales_YM}
\end{table}

\section{Fixing the scale for large $N_c$ gauge theories with adjoint fermions}
Having presented the methodology in the previous sections, we are now going to describe the results obtained by using it on large-$N_c$ gauge theories involving $N_f$ degenerate flavours of adjoint Wilson fermions. The configurations that we have used were generated earlier for other scientific purposes. Reflecting this different origin, our results allow different types of analysis. On one hand for the case of Yang-Mills ($N_f=0$) and Susy Yang-Mills theory ($N_f=\frac{1}{2}$) we have results for a large number of values of $b=1/\lambda$, allowing a determination of the beta function and a study of asymptotic scaling. This will be presented in the next subsection. On the other hand,  for two values of $b$ (0.35 and 0.36) we  have  configurations for $N_f=1/2$, $N_f=1$ and $N_f=2$ covering a wide range of  hopping parameter $\kappa$ values. This would allow us to study the quark mass and $N_f$ dependence of the scale. The technical challenges and main conclusions of this analysis will be presented in the final subsection.  

\subsection{ Scaling analysis for pure and SUSY Yang-Mills  }

In the case of pure gauge theory, we have at our disposal several ensembles at 7 different values of the inverse 't Hooft coupling $b=0.355,0.36,0.365,0.37,0.375,0.38,0.385$. In our analysis, we always use the biggest value of $N_c$ available, mainly $N_c=625$, and $N_c=841$.
For these large values of $N_c$, as explained in the previous section, we do not expect the results to be affected by finite volume effects. We report the values of these scales in Tab.~\ref{tab:achiym}.

A first observation is that our lattice spacing computed in units of $\sqrt{8t_1}$ is perfectly compatible within errors with the scale dependence derived from the string tension \cite{Gonzalez-Arroyo:2012euf}. In fact, the dimensionless quantity defined as $R=\sqrt{8t_1\sigma}$ results to be compatible with a constant value $0.6589(34)$ for all values of the coupling. This comparison with the string tension is a remarkable confirmation of the validity of scaling for a wide range of 't Hooft couplings. Furthermore, our new determination of the scale
has smaller errors than the one previously extracted using the string tension.


 \begin{table}
    \centering
    \begin{tabular}{cccc}
    \toprule
        $b$ & $N_c$ & $\frac{a}{\sqrt{8t_1}}$ & $a\sqrt{\sigma}$\\
    \midrule
        0.355  & 841 &  0.3736(21) &  0.2410(30) \\ 
        0.360  &  625 & 0.3159(16) &  0.2058(25) \\
        0.365  &  625 & 0.27137(88) &  0.1784(17) \\
        0.370  &  625 & 0.2353(31) &  0.1573(19) \\
        0.375  &  625 & 0.2055(14) &  0.1361(17) \\
        0.380  &  625 & 0.1793(13) &  0.1191(17) \\
        0.385  &  841 & 0.1585(15) &  0.1049(11) \\
    \bottomrule
    \end{tabular}
    \caption{Lattice scale for $N_f=0$}
    \label{tab:achiym}
\end{table}

On the other hand, in the case of $N_f=\frac{1}{2}$, we only have configurations for $N_c=289,361$ for 4 values of the coupling $b=0.34,0.345,0.35,0.36$. In \cite{Butti:2022sgy}, after performing the scale setting of these configurations, we performed a linear extrapolation to obtain the value of the scale in the massless-gluino limit, which corresponds to the supersymmetric theory \cite{Kaplan:1983sk,Curci:1986sm}.

\begin{figure}
    \centering
    \includegraphics[scale=0.5]{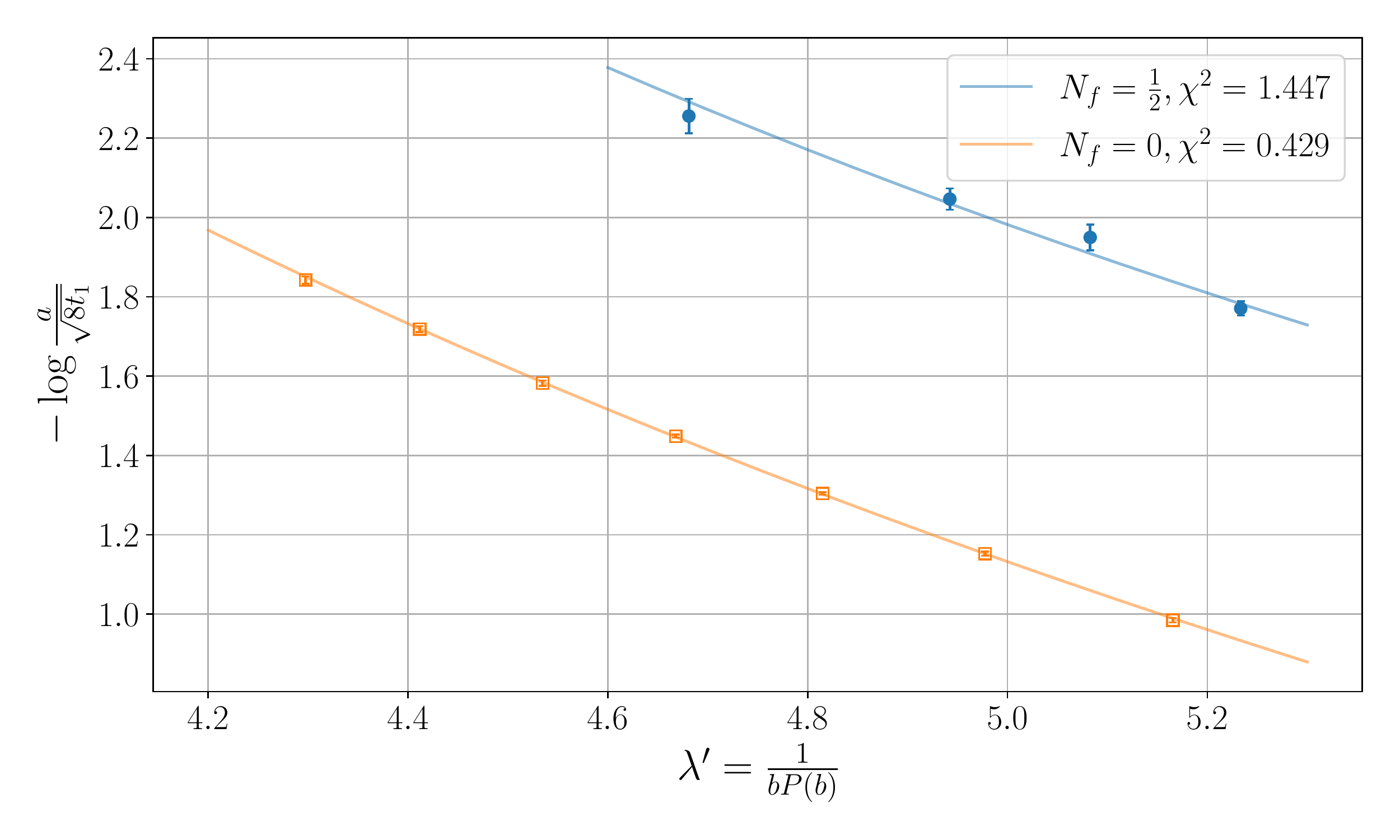}
    \caption{Logarithm of the lattice spacing in units of $\sqrt{8t_1}$ as a function of the improved coupling $\lambda'$. The dependence is given by the $\beta$-function. Data for $N_f=\frac{1}{2}$ are taken from \cite{Butti:2022sgy}. The solid lines and the $\hat{\chi}^2$ in the labels refer to a fit to Eq.~\eqref{eqn:loga}.}
    \label{fig:beta}
\end{figure}

Having the lattice spacing at several values of $b$ for both $N_f=0$ and $N_f=\frac{1}{2}$, we can extract from them the bare $\beta$-function of the theory. 
In order to give a prediction to confront our data, we employ a parameterization of the $\beta$-function which is well suited for the integration
\begin{equation}\label{eqn:NSZV}
    \beta(\lambda) = -\frac{b_0 \lambda^2}{1-\frac{b_1}{b_0}\lambda - \sum_{n\geq 3}c_n \lambda^{n-1}},\quad \text{ with } b_0 = -\frac{4N_f-11}{24\pi^2}\,\text{ and } b_1 = -\frac{16N_f-17}{192\pi^4}.
\end{equation}
In this case, by integration this leads to
\begin{equation}\label{eqn:loga}
    -\log{\frac{a}{\sqrt{8t_1}}} = \log\Lambda + \frac{1}{b_0\lambda} + \frac{b_1}{b_0^2}\log\lambda + \frac{c_3}{b_0}\lambda + \order{\lambda^2} ,
\end{equation}
which can be fitted to our data. It is well known that the naive coupling $\lambda$ does not exhibit precocious asymptotic scaling. We, therefore, use a standard improved version of the coupling $\lambda'$, defined as~\cite{Allton:2007py}
\begin{equation}
    \lambda' = \frac{1}{b P(b)},
\end{equation}
where $P$ is the expectation value of the plaquette. In Fig.~\ref{fig:beta} we plot the logarithm of the lattice spacing in units of $\sqrt{8t_1}$ as a function of the improved coupling $\lambda'$ for the cases of SUSY ($N_f=\frac{1}{2}$) and pure Yang-Mills ($N_f=0$) together with a fit to Eq.~\eqref{eqn:loga}, 
leaving $c_3$ and $\log\Lambda$ as free parameters. The fits are indeed very good, with small reduced $\chi^2$, as displayed in the label of the figure.
 
\subsection{Dependence of the scale on the quark mass and the number of flavours}
In this subsection, we analyze for different theories the dependence of the scale on the value of the adjoint fermion mass.  We focus on the case of $b=0.350$ and 0.360 for which we have data in a large range of fermion masses at all values of $N_f$.  Having those at our disposal, we can attempt a quantitative study of the behaviour of the corresponding theories, trying to highlight the expected substantial differences.

\begin{figure}
    \centering
    \includegraphics[scale=0.4]{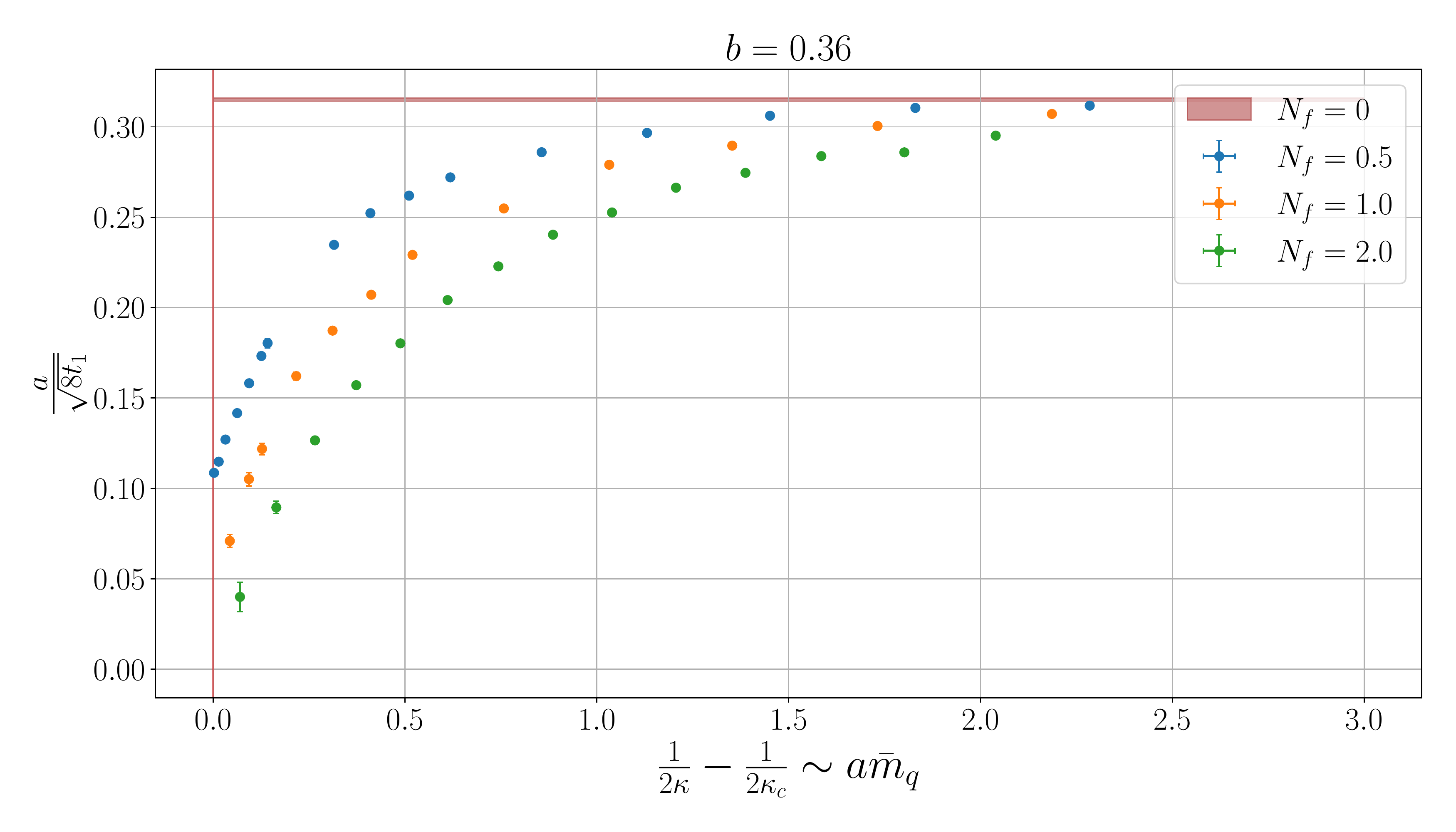}
    \caption{Lattice spacing in units of $\sqrt{8t_1}$ as a function of the bare subtracted quark mass $\frac{1}{2\kappa_a} - \frac{1}{2\kappa_c}$ for $N_f=0,\frac{1}{2},1,2$. The Yang-Mills value is represented as a red band whose width corresponds to the error.}
    \label{fig:a_vs_mq}
\end{figure}

Our results for the lattice spacing at $b=0.36$ as a function of the \textit{subtracted (bare) quark mass}, $a\bar{m}_q \equiv 1/(2\kappa) - 1/(2\kappa_c) $ are displayed in Fig.~\ref{fig:a_vs_mq}. The critical value of the hopping parameter has been extracted by tuning the PCAC mass to zero, following the methodology explained in \cite{Butti:2022sgy}; we obtain $\kappa_c=0.18418(2)$, 0.1777(1) and 0.1698(1), for $b=0.360$ and $N_f=\frac{1}{2}$, 1 and 2 respectively.

\begin{figure}
    \centering
    \includegraphics[scale=0.25]{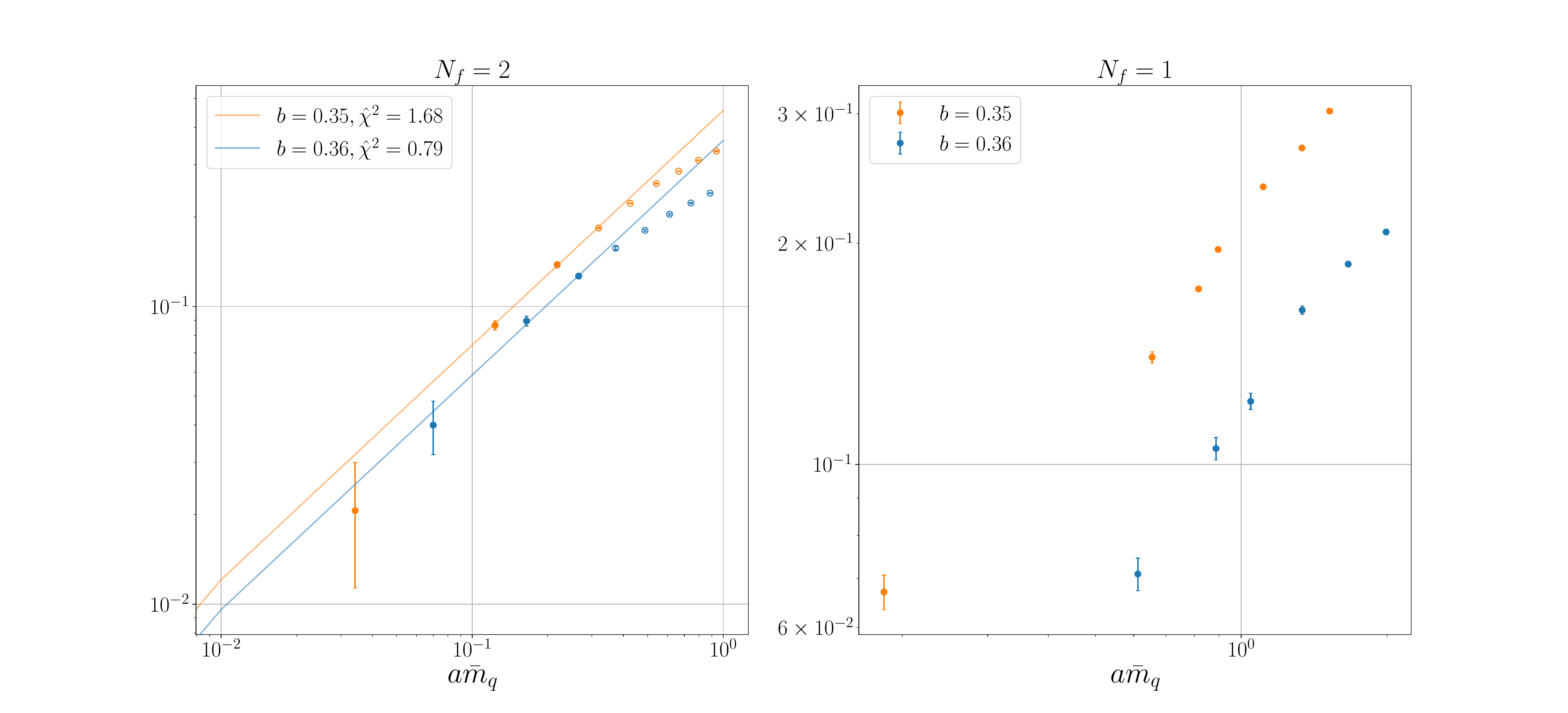}
    \caption{Light sector of the lattice scale. \textit{Left panel:} Lattice scale for $N_f=2$ at $b=0.35,0.36$. Solid points are fitted to $p_1(a\bar{m}_q)^{\frac{1}{1+0.269}}$, the corresponding reduced $\chi^2$ are reported in the label. \textit{Right panel:} Lattice scale for $N_f=1$ at $b=0.35,0.36$.}
    \label{fig:light}
\end{figure}

The plot shows a smooth dependence of the lattice scale on the fermion mass. As expected from the decoupling of fermions in the heavy mass limit, the curves approach a unique value irrespective of the number of flavours and approximate the red band in the figure which represents the value of the scale extracted on pure gauge configurations. 

In the light fermion sector, our results approach different theories. The $N_f=\frac{1}{2}$ case was studied in detail in a former publication~\cite{Butti:2022sgy}. We observed that $a/\sqrt{8 t_1}$ shows a linear behaviour in $\sqrt{8t_1} m_{\rm pcac}$, leading to the scales in the chiral limit given in ref.~\cite{Butti:2022sgy}, which have been used in the determination of the $\beta-$function of $N_f=\frac{1}{2}$ presented in the previous subsection.

The case of $N_f=2$, on the other hand, is known to behave differently. 
In fact, the theory with 2 dynamical adjoint fermions is believed to be conformal in the chiral limit. Indeed, in our previous study \cite{GarciaPerez:2015rda}, we showed compatibility with conformality and extracted the value of the mass anomalous dimension $\gamma^*=0.269(2)(50)$. Given the lack of enough precise data in the small mass region, we can only test the compatibility of our results with the previous ones. Conformality implies that every scale in the theory should behave like $(a\bar{m}_q)^{\frac{1}{1+\gamma^*}}$ as the fermion mass is tuned to zero. This dependence has been fitted to our data fixing the anomalous dimension and leaving free the overall coefficient, showing an agreement within errors as displayed in Fig. ~\ref{fig:light} in a log-log plot.\footnote{An early attempt to extract the anomalous dimension using the dependence on the scale can be found in \cite{Gonzalez-Arroyo:2012vhw}.}

Finally, the case of $N_f=1$ has also been argued to be conformal in a series of recent works~\cite{Athenodorou:2021wom}. The fermion mass dependence of the scale is also displayed in fig.~\ref{fig:light}; as in the $N_f=2$ case,a conclusion cannot be drawn based in our data and more precise results at lighter masses would be required to test this hypothesis.

\acknowledgments
This work is partially supported by grant PGC2018-094857-B-I00 funded by MCIN/AEI/ 10.13039/501100011033 and by “ERDF A way of making Europe”, and by the Spanish Re- search Agency (Agencia Estatal de Investigacion) through grants IFT Centro de Excelencia Severo Ochoa SEV-2016-0597 and No CEX2020-001007-S, funded by MCIN/AEI/10.13039 /501100011033. We also acknowledge support from the project H2020-MSCAITN-2018- 813942 (EuroPLEx) and the EU Horizon 2020 research and innovation programme, STRONG-2020 project, under grant agreement No 824093.
M. O. is supported by JSPS KAKENHI Grant Number 21K03576. K.-I.I. is supported by MEXT as “Program for Promoting Researches on the Supercomputer Fugaku” (Simulation for basic science: from fundamental laws of particles to creation of nuclei, JPMXP1020200105) and JICFuS. This work used computational resources of SX-ACE (Osaka U.) and Oakbridge-CX (U. of Tokyo) through the HPCI System Research Project (Project ID:  hp220011, hp210027, hp200027, hp190004). and Subsystem B of ITO system (Kyushu U.). We acknowledge the use of the Hydra cluster at IFT and HPC resources at CESGA (Supercomputing Centre of Galicia).

\bibliographystyle{JHEP}
\bibliography{bibliography}

\end{document}